# Myths About Congestion Management in High-Speed Networks [1]

R. Jain

Digital Equipment Corp, 550 King St. (LKG 1-2/A19), Littleton, MA 01460, U.S.A
Internet: Jain@Erlang.enet.DEC.Com

## Abstract

Weaknesses in several recently proposed ideas about congestion control and avoidance in high-speed networks are identified. Both sides of the debate concerning prior-reservation of resources versus walk-in service, open-loop control versus feedback control, rate control versus window control, and router-based control versus source-based control are presented. The circumstances under which backpressure is useful or not useful are discussed, and it is argued that a single congestion scheme is not sufficient, but that a combination of several schemes is required for complete congestion management in a network.

# Introduction

With the advent of gigabits per second links and networks, the interest in congestion management [2] is increasing. This is because high-speed links have to coexist with low speed networks of the past that will continue to be used for quite some time. The resulting mismatch of speeds is bound to create congestion. The range of link speeds

---

[1] An earlier version of this paper was presented at the IFIP TC6 4th Conference on Information Networks and Data Communication, Helsinki, Finland, March 1992.

[2] The term "congestion control" is generally used for congestion recovery mechanisms. The term "congestion management" is used here to include both congestion recovery, as well as congestion avoidance.

in networks is continuously expanding. Today, we manage networks with link speeds of 9.6 kbps to 100 Mbps. Tomorrow, we will need to manage networks with link speeds of 9.6 kbps to several Gbps. This increasing heterogeneity is aggravating the congestion problem

Congestion management in high-speed networks is currently a hotly debated topic. Contradictory beliefs exist on most issues. Most of the debate is of a religious nature in the sense that believers on one side are not willing to consider the merits of arguments on the other side.

This paper takes a somewhat devilish view, pointing out weaknesses in several of the ideas that have been recently proposed. This view should not be taken to mean that these ideas are not worth of pursuit. The purpose is to present both sides of issues, permitting an objective comparison of the alternatives.

This paper begins with several old myths about congestion that were presented in an earlier paper [1] and is followed with a number of new myths. The new myths are based on a number of assumptions about the high-speed network. The remainder of the paper discusses the validity of these assumptions.

## Old Myths

An overview of the congestion problem and the factors that affect its design along was presented in Jain [1]. In that paper, the following statements were shown to be false:

1. Congestion is caused by a shortage of buffer space. The problem will be solved when the cost for memory becomes cheap enough to allow infinitely Large memories.

2. Congestion is caused by slow links. The problem will be solved when high-speed links become available.

3. Congestion is caused by slow processors. The problem will be solved when processor speed is improved.

4. If not one, then all of the above developments will eliminate the congestion problem

These old myths are based on the belief that as resources become less expensive, the problem of congestion will automatically be solved. It was shown in [1] that increasing memory sizes, processor speeds, and link bandwidths has actually aggravated the congestion problem. Proper inclusion of congestion management and avoidance mechanisms in protocol design is more important today than ever before.

# New Myths

A number of new myths have been observed since that earlier paper was published. The following claims have been heard at various high-speed networking workshops:

1. The traffic on high-speed networks will be primarily video-like (steady and predictable). Therefore, prior-reservation of resources is required in place of a datagram service.

2. Large quantities of data in the pipe at high speed requires an open-loop control scheme instead of a feedback scheme.

3. Rate control must be used in place of current window controls.

4. Source-based control schemes, which require sources to be informed of the congestion, are too slow for high-speed networks. Router-based controls must be used instead.

5. Backpressure is the ideal congestion control scheme for high-speed networks since it provides an immediate relief.

6. A single congestion management scheme is sufficient.

These statements are not always true. As shown in this paper, they may be true under specific circumstances, but are false under other circumstances. However, the circumstances have little relationship with the speed of the network. Given a specific set of circumstances, the statement will be true (or false) for a low-speed, as well as for a high-speed network.

The key issue in the design or selection of a congestion management scheme is the traffic pattern, and traffic patterns are dependent upon the application. This topic is covered in the next section.

## Traffic Patterns on High-Speed Networks

The first issue that helps decide the traffic pattern is the use of high-speed links. Could they be used as a backbone to interconnect slower subnets (Figure 1a) or could they be used as subnets interconnected via slower links (Figure 1b)? As it stands today, high-speed technology is used in local-area networks (LANs), which are interconnected via a slow wide-area networks (WANs). The argument that favors this setup is that network traffic is highly local. The traffic traveling between the subnetworks is considerably less than the traffic on the subnetwork itself.

To appreciate the opposing view, consider automobile traffic on the highway. High-speed roads (highway) are used to interconnect slow-speed roads (city lanes). This is so in spite of the fact that automobile traffic is primarily local.

The slow speed of the WANs is the result of the unavailability of high speed WANs. When available, high-speed links will replace current WAN links. For example, FDDI backbones will be used to interconnect 10-Mbps Ethernet LANs and 16-Mbps token rings. Gigabit per second (Gbps) links will be used as a backbone for FDDI LANs and so on. Configuration shown in Figure 1a has severe congestion problems at the entry to the backbone. The congestion results because the nodes on subnets are capable of high-speed communication and when two nodes of different subnetworks communicate, the traffic coming into the backbone needs to be processed at a high speed.

There is also an economic reason for high-speed backbones. High-speed links are more expensive than low-speed links. Since expensive resources are generally shared, the higher the expense, the greater is the sharing. Thus, *high-speed links will be shared by a large number of nodes on the low-speed subnets.*

There are three implications as a result of this increased sharing of high-speed links. First, the speeds of individual sources do not have to be in gigabits per second, although switches, bridges, routers, gateways, and other shared resources have to be capable of handling Gbps.

Second, a greater variety of applications sharing the network implies that the networks will have to satisfy a variety of performance criteria. Some applications, such as voice and video, are delay-sensitive, but loss-insensitive. Others, such as file transfer and electronic mail, are delay-insensitive, but loss-sensitive. Still others, such as interactive graphics or interactive computing applications, are delay-sensitive and loss-sensitive. Any scheme that distinguishes between the sources (for example, queueing and service strategies), but treats all applications identically will not be helpful. *Any attempt to achieve fairness under overload would have to allow for differing application requirements at these sources.*

Third, the telecommunication and computer networks of today have primarily been designed in isolation. In telecommunication networks, bandwidth has a price, while in the computer networks, only interfaces do. The telecommunication networks have been designed for applications with a predictable bandwidth requirement and the users are charged for the bandwidth. The computer networks have been designed for applications that can share any available bandwidth and the users pay only the price of the interfaces (adapters). Data on telecommunication networks is treated like voice with reserved bandwidth and no strict reliability (error) or loss requirements. Similarly, voice on computer networks is treated like data packets, except possibly at a high priority.

Due to the increased cost of high-speed networks, the separation of applications is not cost effective. Traffic on the high-speed links will be a mixture of several applications. Designing a high-speed link with just one application in mind, such as file transfer or video distribution, is not prudent. The higher the speed, the more heterogeneous the traffic. Future networks will carry data, voice, video, and other multimedia traffic. The design of congestion management techniques should accomodate characteristics of all these different applications.

## Window or Rate Control?

Flow controls utilizing window mechanisms are used in almost all existing computer networking archetectures, including TCP/IP, DNA, OSI, and SNA. A rate-based resource allocation is common in telecommunication networks where each connection has a specific bandwidth assignment. Recently, several protocols with rate-based flow control have been proposed for computer networks. In this approach, the destination node specifies the maximum rate (number of packets over a given time) at which the sources can send packets.

Some have argued that in future rate-based controls will replace window-based controls. This is justifiable since memory will no longer be the bottleneck. Instead, the processors, links, and storage devices will be the bottlenecks. These latter resources are rate limited in the sense that they cannot sustain bits or packets arriving at a rate faster than their capacity. Memory was previously the bottleneck and it was count limited in the sense that it could not sustain more than a certain number of packets (or bits) regardless of how fast or slow they came in. The window-based flow control schemes originated from the desire to keep the bottleneck (memory) from overflowing. The second problem with window-based controls is that in some implementations all packets of a window can be transmitted back-to-back, resulting in bursty traffic. The third argument for rate-based controls is that much of the high-speed traffic will be stream oriented (such as voice or video), requiring a guarantee based on rate rather than count. This is unlike data traffic that generally requires sending a certain number of packets.

Unfortunately, there are several misconceptions about rate-based controls. First, it is generally not understood that *specifying the rate requires at least two quantities*: the number of packets $n$ over a period $T$. Given a permissible rate of $n/T$ packets per second, there are several possible values of $n$ and $T$ that will result in the given rate. However, not all possible combinations may be acceptable. A switch that can process 1 packet per millisecond may not be able to handle 5 packets arriving back-to-back every 5 milliseconds (see Figure 2). Therfore, all rate-based schemes, including the popular leaky bucket [2] and its variations [3,4,5,6], require specifying the burst size and the interburst interval. Analytical models of rate-based controls generally ignore the two parameters and instead model the arrival process with a single parameter $\lambda = n/T$.

Second, it is generally not understood that rate-based controls are hop-by-hop mechanisms since all intermediate systems (routers or bridges) should be made aware of the rate parameters and should enforce them. These controls cannot be enforced with end-to-end mechanisms alone. Without hop-by-hop enforcement, several bursts may be combined into a single burst by the intermediate systems. For example, a bridge that is not aware of the rate parameters may change a 1 packet per millisecond stream into a 5 packets (back-to-back) per 5 millisecond stream. The destination (or whatever specified the control) may not be able to process or accept the altered stream (Imagine viewing a video with arrivals more bursty than those allowed by receiver buffering.)

The essence of this argument is that *rate-based controls require a connection-oriented approach* since the parameters $n$ and $T$ must be agreed upon by all intermediate systems along the way. Implementing rate-based controls in a connectionless network is difficult. In a connection-oriented network, if there are bridges that are not involved in flow-control decisions, but get congested, rate-based controls are difficult to enforce. Window-based controls, on the other hand, can be applied end-to-end, hop-by-hop, or using both. In the end-to-end version, the intermediate systems do not have to be informed about the windowsize set by the destination.

It should be clear from the above discussion that a rate-based admission control alone is not sufficient (on entry to the network). Rate controls have to be enforced at every node in the network. Most discussions on rate-based control, such as a leaky bucket scheme and its variations, are limited to admission control. These may work on networks with small number of hops but will need to be supplemented by switch/router mechanisms for large networks. Recently, Golestani [7] has proposed a stop-and-go service policy that allows rate enforcement at intermediate nodes.

Third, *with dynamic rate-based controls there is a possibility of significant increase in packet queue when the total input rate is close to the capacity.* Leland [8] shows that rate-based controls result in a higher total delay (including delay at the source queue, as well as in the network) and a higher loss rate than window-based controls at loads in excess of 60% of the capacity. Depending upon the feedback and control delay, the queue lengths may increase to several thousand packets. This is particularly true if the feedback is delayed or lost. If the memory capacity is available, storing large numbers of packets leads to unacceptably high delays. In the extreme case of infinite feedback delay, a rate based control will result in the input continuing unchanged forever. With window based controls, input to the network stops automatically as the windows are exhausted due to increased feedback delay. In other words, *window schemes are inherently close-loop, while rate-based schemes are inherently open-loop.*

The need to keep queues within a reasonable bound requires that the rate-based controls be modified to become closed-loop. One way, for example, is to supplement the rate-based control with a large window limit. This window limit will generally not be exercised. But, if a rate mismatch occurs and queues build up, the sources will run out of their window quotas and will stop further injection of traffic until new acknowledgement from the destination opens the window again. Windowsizes in current systems are 1 to 32 packets. When used as a backup to a rate-based control, the windowsizes would be one or two orders of magnitude higher. Such a combination has recently been proposed Keshav [9].

This discussion of window-based versus rate-based controls is summarized in Table 1.

## Open-Loop or Feedback?

Given a high-speed network in the gigabit per second range and given today's packet sizes, it is generally true that the propagation delay (the time for the first bit to travel the network) is considerably higher than the packet transmit time (the time between the first and the last bit of the packet). As the link speed increases, the number of bits traveling in the link (also called "in the pipe") increases. Therefore, the number of packets that may be in the pipe increases. A simple calculation shows that a coast-to-coast (3000 mile) 1-Gbps fiber link can accommodate approximately 24 Mbits of data. Given an average packet size of 512 bytes[3] approximately 6000 packets can be in the pipe.

Many of the old congestion management schemes are close-loop schemes in the sense that congested resources send a feedback signal to the source of traffic, which then adjusts the traffic level. It has been argued that such schemes are too slow since by the time a source gets the feedback and reacts to it, several thousand packets may have been lost. This has led to the development of several open-l[4] approaches that do not require feedback. Router-based controls, prior-reservation, and backpressure are examples of open-loop schemes. The relative merits of these schemes are discussed in the next few sections.

---

[3] In todays networks, average size is of the order of 128 to 256 bytes.

[4] Since a loop is always closed, some say that open-loop is an oxymoron.

Table 1: Window-based Control versus Rate-based Control

|  | Window-based | Rate-based |
|---|---|---|
| Control | Window ($W$) | Number of packets ($n$), and Time interval ($T$) |
| Effective rate | $\frac{\text{Window}}{\text{Round-trip delay}}$ | $\frac{n}{T}$ |
| Required if | Memory is the bottleneck | Processor, link, or other devices are bottlenecks |
| Maximum queue length | Limited to sum of windows | No limit |
| Burstiness | Results in bursty traffic | Not bursty at the source |
| Control span | End-to-end, hop-by-hop, or both | Hop-by-hop |
| Network layer | Connectionless or connection-oriented | Connection-oriented |

# Router-Based or Source-Based Controls?

The routers, bridges, switches, multiplexors, and gateways form the core of computer networks that are shared by several end-systems. Here, the term router is used to denote all such intermediate nodes and the term source denotes all end-systems. In the past, many congestion management schemes proposed that the routers send a feedback signal to the sources, which will initiate remedial control action increasing or decreasing the load. Examples of source-based controls include the slow-start [10], CUTE [11], the DECbit [12], and the Q-bit [13] schemes.

There are several arguments against source-based controls. First, these controls have a significant delay time between sensing the congestion and taking the remedial action. If the session lengths are short, the feedback may arrive after the source has finished transmitting all of its data. Second, sources may or may not cooperate and follow the directions from the network. For example, it is possible for a source to reduce its retransmission interval under high-loss conditions, achieving a higher success rate. Third, in some schemes the feedback requires that additional packets be injected into the network.

The router-based controls do not suffer from these problems since they evenly distribute their resources without relying on the sources. Shenker [14] has proven that router-based controls are necessary for fairness and that source-based controls can achieve efficiency, but may not always be fair.

Examples of router-based controls are random drop policy [15], fair queueing [16], and backpressure. While source-based controls have been successfully used in private networks where sources, as well as the network, are owned by the same organization, such controls need to be supplemented with router-based controls in a public network environment where the routers are owned by a telecommunications company, which may have no control over the sources. Source-based controls generally use network layer protocols to transmit the feedback (if explicit) and transport layer protocols to reduce traffic. Therefore, the presence of multiple protocols at these layers makes the implementation of source-based controls unfair. For example, in a network with both OSI/TP4 and TCP/IP transports in use, upon packet loss the sources using OSI/TP4 may reduce traffic to a different level than those using the TCP/IP.

The key problem with router-based controls is that *they introduce complexity in the routers that are shared resources*. In fact, in many of the proposed schemes the complexity is proportional to the number of sources[5] sharing the router. This number increases as the network link speeds increases. (A gigabit per second link can be shared by many more sources than a 300-baud link.)

In our experience with the implementation of the DECbit scheme, we found that implementors have no problem with the transport layer (source) part of the scheme, but some are hesitant to implement the network layer (router) part of the scheme, since it introduces 10 to 12 instructions per packet in the forwarding path. As the network speeds increase, the number of processor cycles per packet rapidly decreases. Any scheme that introduces additional cycles in the router code is undesirable. Implicit feedback schemes, such as those using delay [17, 18] as the signal, are more desirable. Another interesting proposal is to send one or more special coded packets so that in the event of a packet loss, information can be reconstructed at the destination without a retransmission [19].

With source-based controls, the source complexity can be different at different sources. Sources capable of generating high loads may have more sophisticated controls, than those generating low loads (for example, interactive traffic). The latter may have simple static controls, such as fixed small windows.

A basic argument against router-based controls is that *unless sources reduce their traffic, the congested state will continue*. This is not a problem for short overload spikes but if the congestion lasts long, the load may spread over adjoining routers and links, and eventually spread all the way to the sources. Therefore, router-based controls are good for enforcing fairness under overload, while source-based controls are required for sustained overload.

---
[5] This assumes that all applications at a source are treated identically. Requiring application level fairness would introduce even more complexity.

Many recent proposals have argued in favor of dropping data packets on overload, as a desirable congestion control scheme at high speed. For some applications, for example, voice and video, loosing a few packets every now and then is acceptable. For most data applications, every packet is as important as the other.[6] Every lost data packet has to be recovered by subsequent retransmission of that packet along with several others. This recovery is generally an expensive operation at any speed. For higher speed networks, the number of packets to be retransmitted is larger due to a larger pipe size. Therefore, there is a need to consider congestion avoidance schemes that act before dropping packets becomes necessary.

The source disobedience problem in public networks can be overcome by putting the controls at network access points (DCEs). The routers (or switches) in the interior of the network would send the feedback to the routers at the network access point which would accept or reject the traffic from individual sources using a fair allocation scheme.

To summarize, in the router-based versus source-based debate, *router-based controls are required for fairness and work under short-duration overloads. While source-based controls are required for longer overloads.* If the packet processing speed is important, one would argue for simplicity in the routers. However, if source disobedience is the issue, then one has to use router-based controls regardless of the speed of the network. Using network access controls is also a possibility. The key points of this debate are summarized in Table 2.

---

[6] Imagine loosing a few digits of a bank check to you.

Table 2: Router-based versus Source-based Controls

|  | Router-based | Source-based |
|---|---|---|
| Examples | Randomdrop | Dynamic window |
|  | Fair queueing | Slow-start |
|  | Backpressure | DECbit |
| Delay | None | Feedback delay |
| Feedback overhead | None | Feedback messages or bits |
| Overhead in | Routers | Sources |
| Required if | No control over sources | Longer overloads |
| Fairness | Achievable | Not guaranteed |
| Overload duration | Short | Greater than feedback delay |

# Backpressure

Backpressure has been reported as the key (or only) congestion control mechanism in several recent proposals [20, 21]. Backpressure is a form of hop-by-hop, on/off flow control. Congested routers send a "transmission-off" (X-off) signal to neighboring routers (or sources) and stop accepting further packets until their queues reduce. When the load reduces, a "X-on" signal is sent and packet flow resumes.

Backpressure is a datalink-level mechanism. A datalink-level mechanism has a shorter feedback loop than the transport-level mechanism. Therefore, it works well if the overload is short-lived. During this short period, the resources of neighboring routers are used to sustain the overload. The backpressure has the same effect as increasing the number of buffers. However, if the overload is of long duration, backpressure may bring the whole network to a standstill. Unless sources are informed to reduce their traffic, the traffic keeps entering the network resulting in a standstill.

For long duration overload, backpressure is more effective in small networks than in networks with larger diameters. This is because in small networks, sources are close to routers and the backpressure signal reaches the sources quickly. It is important to understand this while doing simulation analysis, since the results based on small diameter networks may not apply to larger networks.

Backpressure is also unfair in the sense that traffic not traveling the congested resources is adversely affected. This is shown in Figure 3. The congestion at router R, spreads to several links (indicated by thick lines). The flow A, which does not use router R, is affected whenever flow B increases its load.

Maxemchuk and Zarki [22] argue that the sharing of high speed links by a large number of sources will, by the law of large numbers, result in smooth traffic patterns. This in turn implies that short-duration overloads (short traffic peaks) are less likely in high-speed networks.

To conclude, *backpressure should only be used for short-duration overloads* after which the pressure should be removed. For long-duration overloads, backpressure should be supplemented with a transport-level or network access level control scheme.

## Prior-Reservation or Walk-in?

Prior-reservation means that sources have to reserve the required resources at connection setup. Prior-reservation in high-speed networks is justified using two arguments. First, much of the traffic is steady voice or video traffic, in which case the required resources are known at the connection setup time. Second, the data traffic is so short-lived that by the time a feedback arrives at the source, the source has probably finished transmitting. For example, sending a facsimile may take only a few milliseconds on a gigabit link, while the feedback may take several tens of milliseconds to travel.

Network users prefer reservation if they want bandwidth or delay guarantees that are difficult to achieve with walk-in service. Reservations also make resource management easier since the demands and capacities are known in advance. With walk-in service the resource management problem is dynamic and rather difficult.

A key disadvantage of reservations is that any resource that is reserved, but not used is wasted. Therefore, while it is good for steady predictable traffic, it is not a good choice for bursty or unpredictable traffic. Distributed systems, such as clusters, require a fast communication mechanism between various processors or processors and memory/storage devices. Currently, these systems use high-speed buses (bandwidth of several hundred megabits per second), but are limited to a few kilometers in extent. In the future, gigabit per second networks with larger extents will permit these systems to cover greater distances. However, the traffic for such systems will continue to be bursty, unpredictable, and unamenable for prior-reservation.

There is some setup overhead incurred in reserving resources. This overhead is not justifiable if the session duration is short as is the case in some computer applications, such as remote procedure calls.

The walk-in schemes do not require maintainance of any state and, therefore, are ideally suited for highly dynamic environments. On the other hand, *reservation schemes are not suitable for highly dynamic environments.*

The reservation versus walk-in debate is summarized in Table 3. The conclusion is that the choice is independent of the speed of the network. *Reservation is good for long, steady sessions, while walk-in service is required for short bursty traffic.* We expect to see both types of traffic in high-speed networks; therefore, networks providing only one type of resource management will not be successful. A few combinations have already been proposed. For example, in one proposal the first packet of a train results in a medium term reservation of resources for the entire train. The reserved resources are deallocated at the end of the packet train. For long trains, prior reservation using a separate set-up packet may be used to reduce the possibility of packet loss due to unavailable resources.

## One Scheme or Many?

Proponents of congestion management schemes claim that their scheme is better than all existing ones and that theirs is all that is required, for example, backpressure alone or admission control alone is sufficient. This unfortunately is a myth. The type of scheme needed is dependent upon the duration of the overload. For example, transport-level dynamic window schemes require several round-trip delays to be effective. If the congestion lasts less than the round-trip delay, they will have no effect other than to cause a source to reduce traffic even though the congestion has disappeared. It is well known that datalink level schemes, such as backpressure, are more effective for short duration overloads. However, it is a lesser known fact that the opposite is also true: *most datalink level schemes are not effective for long duration congestion.* As a general rule of thumb, *the longer the duration, the higher the layer at which control should be exercised.* For example, if the congestion is permanent, the installation of additional links or high-speed links is required. If the congestion lasts for a session duration, a session level control (such as, a busy signal) is more appropriate. If the congestion lasts for several round trip delays, transport level controls (with feedback from the network layer) are more effective. If the congestion is of a short duration, network level controls (various router-based or admission controls) and datalink level controls (backpressure) should be used. Since every network can have overloads of all durations, every network needs a combination of controls at various levels. No one scheme can solve all congestion problems.

Table 3: Reservation versus Walk-in

|  | Reservation | Walk-in |
|---|---|---|
| Guarantees | Guaranteed bandwidth and/or delay | Varying bandwidth or delay |
| Resource management | Easy | Difficult |
| Unused resources | Wasted | Can be used by other sources |
| Good for | Steady traffic (Voice/Video) | Bursty traffic. (Data) |
| Setup | Setup required $\Rightarrow$ Good for long sessions | No setup required $\Rightarrow$ Good for short sessions |
| State | More state $\Rightarrow$ Less dynamic | No state $\Rightarrow$ More dynamic |

An example of a combined approach is given in a proposal by Eckberg, Luan, and Lucantoni [23]. This proposal requires a leaky bucket admission control for normal operation, a source-based control for packet loss, and a session-denial for longer term congestion.

Another related issue is that of multiple competing schemes at the same level. In standards committee meetings, the lack of agreements on competing but essentially similar proposals is often resolved by allowing the option of using any (zero or one) of the proposed schemes. In most cases, this results in unfair and uncontrolled networks. For example, differing increase and decrease algorithms at the transport layer used by different vendor nodes may give unfair advantage to some nodes. Unlike other parts of networking architectures, congestion control deals with shared resources and *it is better to have one rule for all players than to let the players choose the rule.*

## Summary

The introduction of high-speed links is causing greater heterogeneity in computer networks, therefore making them more susceptible to congestion.

Several ideas have been proposed for handling congestion in high-speed networks. The principal ones are rate control, open-loop control, prior-reservation, and router-based controls. These proposals and their opposites, such as window-based control, feedback control, walk-in service, and source-based controls have been compared and arguments for and against each case presented.

High-speed links of the future will be shared by many more sources and applications than the links of today. As a result, the higher the speed, the more heterogeneous the traffic. Designing a high-speed link for data traffic or voice/video traffic alone is not prudent. Rate-based controls must be exercised at each loop. These controls may result in long queues unless they are backed-up by large, window-based controls. Router-based controls are required for fairness and work under short duration overload. For long duration overload, source-based controls are required to reduce input traffic. The same applies to backpressure. If backpressure is used it should be limited to short duration and be supplemented with higher level controls. Reservation is good for long, steady sessions, but not for data traffic, which is bursty, unpredictable, and dynamic.

A complete congestion management strategy should include several congestion controls and avoidance schemes that work at different levels of protocols and can handle congestion of varying duration. In general, the longer the duration, the higher the protocol layer at which control should be exercised. Any one layer, such as datalink (backpressure) or routing (queueing/service strategies), cannot handle all congestion problems.

Figure Captions

Figure 1: Two possible applications of high speed networks. First one has severe congestion problem

Figure 2: Rate-control requires specifying the number of packets and a time interval. One packet per millisecond is different from 5 packets per 5 milliseconds. Each vertical arrow represents a packet arrival.

Figure 3: Backpressure if sustained for long duration can result in congestion spreading throughout the network. Flows that are not using the congested resources may also be affected.